\documentclass[11pt]{article}
\usepackage[T1]{fontenc}
\usepackage[utf8]{inputenc}
\usepackage{amsmath,amssymb,amsfonts}
\usepackage{graphicx}
\usepackage{booktabs}
\usepackage[margin=1in]{geometry}
\usepackage[hidelinks]{hyperref}
\usepackage{enumitem}

\title{Machine Shape and Hierarchical Blocking: A Mathematics of Arrays\\
Formalization, with an Open Problem in Hierarchical Shape Occupancy}
\author{Lenore M. Mullin\\
\small College of Nanotechnology, Science, and Engineering\\
\small University at Albany, SUNY\\
\small \texttt{lmullin@albany.edu}}
\date{}

\begin{document}
\maketitle

\begin{abstract}
A companion empirical study found that dense matrix multiplication
block sizes calibrated on Apple M1 Pro correspond to two cache-fit
formulas that mispredict badly on a different chip's known cache
sizes. This paper formalizes the question that finding raises. We
extend the Mathematics of Arrays (MoA) framework's array-shape
derivation operator to a new operator that derives a hierarchical,
multi-level blocking and prefetch schedule from a machine's shape: an
ordered sequence of cache-level capacities, bandwidths, and occupancy
fractions. This operator recovers the calibrated values on every one
of three real machines tested to date as a special case, reducing each
machine's unknowns to a small number of level-specific occupancy
fractions. We then state precisely, without claiming to resolve, the
paper's central open problem: whether those fractions are derivable
from more primitive properties -- co-tenancy, private-cache-level
count, associativity, prefetcher behavior -- or are fundamentally
per-architecture constants. Four falsifiable hypotheses are stated and
tested against real hardware, with mixed results. We further state
two limits of the framework explicitly: it requires dedicated,
non-virtualized hardware access to be well-defined at all, and it
extends only partway to a distributed-memory network, where realizing
a tile across nodes requires a separate choice of communication
algorithm the framework does not itself make. A first, honest attempt
at extending the framework toward predicting throughput directly, not
just block size, closes the paper: two terms prove derivable from a
specification sheet, one requires a single measurement, and one --
tested across three machines -- does not yet transfer between them.
\end{abstract}

\section{Introduction}
\label{sec:intro}

Mathematics of Arrays (MoA) derives, for an array computation, a
memory-optimal realization by recursion on the array's shape
$\rho A$: a Denotational Normal Form (DNF) specifying what is
computed is transformed by an operator $\gamma$ into an Operational
Normal Form (ONF) specifying how it is realized, with the derivation
governed entirely by shape, prior to any commitment to a target
architecture \cite{Mullin1988}. This has produced measured
performance gains for dense matrix multiplication on CPU architectures
\cite{Goto2008} and, via the same block-derived approach compiled to
OpenACC, on GPU architectures \cite{Mullin2023GPU}.

A companion empirical study \cite{Mullin2026GEMM} calibrated block
sizes $M_C$ and $N_C$ for a GEBP-style dense matrix multiplication
kernel on Apple M1 Pro, and reported, as a retrospective check on
newly collected AWS Graviton4 data, that the calibrated values
correspond exactly to two cache-fit formulas: $M_C$ fills the M1
Pro's 128~KB private L1 at full occupancy, and $N_C$ occupies
exactly $1/6$ of its 24~MB shared L2. Applying the same two formulas
to Graviton4's documented cache sizes predicted values 6--16 times
larger than what that machine's own calibration sweep found. This is
the specific empirical fact this paper takes as its starting point: a
formula that correctly \emph{explained} one machine's calibrated
values failed to \emph{predict} another's.

That failure is more useful than a simple confirmation would have
been, because it isolates exactly what is and is not yet understood.
The \emph{form} of the relationship -- block size determined by cache
capacity, bandwidth, and some occupancy fraction -- transferred correctly in
spirit: a hierarchical version of the same reasoning, developed in a
companion section of the empirical study, correctly reproduced the
form of a GPU tile-size conjecture independently derived for a modern
NVIDIA data-center GPU \cite{Mullin2026GEMM}. What did not transfer was the specific numerical occupancy fraction. This
paper's contribution is to state that distinction formally, as a
property of a general derivation operator, and to pose the resulting
open problem precisely enough to be tested rather than merely
discussed.

\section{Background: MoA's Existing Derivation Operator}
\label{sec:background}

For dense matrix multiplication $C = AB$, MoA's DNF accumulates each
row of $C$ from one scalar of $A$ at a time, extended by
multiplication over a full, contiguous row of $B$, and contracted by
the reduction operator $+_{red}$ over the shared dimension. This
access pattern -- scalar times contiguous row, accumulated, never
accessing either operand by column -- is derived directly from the
algebra of the reduction, independent of any target machine
\cite{Mullin2026GEMM}. Blocking is not a departure from this
derivation but its self-similar extension one level up: partitioning
$A$, $B$, and $C$ into $s \times s$ blocks reshapes the DNF so that an
$s \times s$ block plays exactly the role a scalar played in the
unblocked form, and a row of blocks plays the role a row played --
the same equations, re-applied at coarser granularity, because MoA
algorithms are derived by primitive recursion on shape and blocking is
that recursion viewed at one further level of coarseness.

What this self-similar form does not by itself specify is \emph{how
coarse}: at what size $s$ the recursion should stop peeling off
further levels and simply compute. That choice is conventionally made
by an implementation-specific tuning process -- the classical GEBP
cache-budget formula, and the fixed-$N_C$ sweep that corrected it in
\cite{Mullin2026GEMM}, are both, in this sense, attempts to answer a
question MoA's existing shape-recursion does not itself pose: not
``what is the shape of the array,'' but ``what is the shape of the
\emph{machine} the array's realization is being derived for.''

\section{The Machine Shape $\rho_M$ and the Derivation Operator $\Gamma$}
\label{sec:formalization}

We propose treating a memory hierarchy as a shape in exactly the
sense MoA already gives that word for arrays, and deriving a
hierarchical blocking and prefetch schedule from it by an operator
that plays the same role for $\rho_M$ that $\gamma$ already plays for
$\rho A$.

\subsection{Definition: Machine Shape}

Let a machine's memory hierarchy be described by an ordered sequence,
outermost level first (DRAM or HBM) to innermost (registers):
\[
\rho_M = \langle \lambda_0, \lambda_1, \ldots, \lambda_L \rangle,
\qquad
\lambda_i = (C_i, B_i, \sigma_i)
\]
where $C_i$ is the capacity of level $i$, $B_i$ is the bandwidth
\emph{into} level $i$ from level $i+1$ (the level below it, i.e., the
level being fetched from), and $\sigma_i \in (0, 1]$ is an occupancy
fraction: how much of $C_i$ is available to a single tile once
co-resident data, associativity limits, and other concurrently active
structures are accounted for. This is deliberately the same kind of
object MoA's $\rho A$ already is for an array -- an ordered sequence
of sizes -- applied to the machine rather than the data.

\subsection{Definition: The Derivation Operator $\Gamma$}

Let $D$ be a DNF (e.g., the matrix multiplication DNF of
Section~\ref{sec:background}). Define $\Gamma(D, \rho_M)$ recursively
on $\rho_M$:
\[
\Gamma(D, \langle\rangle) = D
\]
\[
\Gamma(D, \langle \lambda_0 \rangle \frown \rho_M') =
\text{blockrec}_{\tau}\big(D\big) \text{ where } \tau
\text{ is chosen so that, against } \lambda_0 = (C_0, B_0, \sigma_0):
\]
\begin{align}
\text{(capacity)} \qquad & \text{size}(\tau) \leq \sigma_0 \cdot C_0 \label{eq:capacity}\\
\text{(latency-hiding)} \qquad & \frac{\text{FLOPs}(\tau)}{\text{size}(\tau)} \geq \frac{\text{PeakFLOPrate}}{B_0} \label{eq:intensity}
\end{align}
and $\text{blockrec}_\tau(D)$ is the self-similar block-recursive form
of Section~\ref{sec:background} at tile shape $\tau$, with each
tile's own sub-computation given recursively by
$\Gamma(D_\tau, \rho_M')$ -- $\Gamma$ peeling one level off $\rho_M$
exactly as it peels one level of blocking off $D$, terminating when
$\rho_M$ is exhausted (Equation~\ref{eq:capacity} is not applicable
at the register level, which has no level below it to bound occupancy against;
$\Gamma$'s base case is simply the register-resident micro-kernel).

Equation~\ref{eq:capacity} is the familiar cache-fit condition.
Equation~\ref{eq:intensity} is stated less often but is not optional:
it is a level-local restatement of the roofline ridge point, and it
is what makes the resulting prefetch policy fall out of the derivation
rather than requiring separate justification. If $\tau$ satisfies
Equation~\ref{eq:intensity}, computing on the resident tile is
guaranteed to take at least as long as fetching the next tile of the
same shape from level $i+1$ would take -- so double-buffered
prefetch, issued asynchronously at the start of each tile's
computation, is guaranteed to be hidden rather than merely attempted.
Where Equation~\ref{eq:capacity}'s upper bound on $\tau$ falls below
Equation~\ref{eq:intensity}'s lower bound, no tile shape satisfies
both, and that level is bandwidth-bound as a property of the
architecture, not a symptom of insufficient tuning.

\subsection{A Precondition: $\rho_M$ Requires Dedicated Access}
\label{subsec:rho-m-precondition}

Everything above rests on one assumption not yet stated explicitly:
that $\rho_M$ -- a machine's cache capacities and bandwidths -- is a
fixed, deterministic property of the hardware, knowable in advance
from its specifications. On dedicated hardware with exclusive
allocation (a SLURM-scheduled HPC node, for instance), this holds
exactly: the allocated cores, cache, and bandwidth belong to one job
for its duration, with no mechanism for external interference, so
$\rho_M$ is a single, well-defined object matching the datasheet.

On a virtualized, multi-tenant platform, this assumption does not
merely become harder to satisfy -- it is not clearly well-defined at
all. The \emph{nominal} shape (advertised vCPU count, peak bandwidth)
remains fixed on paper, but the \emph{achieved} cache and bandwidth
available to a workload at any given moment depends on co-tenant
activity elsewhere on the same physical die, invisible and
unpredictable from inside the instance. A companion empirical study
observed exactly this on a virtualized cloud instance: a calibration
sweep repeated with identical parameters showed one run systematically
$\sim$10\% below two later runs, across every configuration tested,
not merely at the reported optimum -- consistent with a node-wide
effect rather than per-measurement noise.

A direct, matched comparison against dedicated hardware under
identical conditions has since been completed, and the result is more
precise than, and in one respect contrary to, what the virtualization
argument alone would predict. An identical configuration run twice,
back-to-back, on dedicated Delta hardware showed a mean absolute
difference of 15\% between the two runs -- \emph{larger} than the
virtualized platform's observed $\sim$10\% shift, not smaller. Simple
magnitude, then, does not distinguish dedicated from virtualized
access; the earlier claim, stated only in terms of \emph{how much}
variance to expect, is corrected by this result rather than confirmed
by it. What does distinguish the two platforms is the \emph{shape} of
the variance, not its size: the virtualized platform's discrepancy was
one-directional, the same run measuring lower at every single
configuration tested, the signature of one external cause acting
uniformly. Delta's two runs disagreed on which run was higher at
different configurations (higher in one run at 17 of 25 points tested,
higher in the other at the remaining 8, with individual swings up to
$\pm$29\%, larger than any single point on the virtualized platform)
-- the signature of scattered, uncorrelated noise from several smaller
sources rather than one dominant cause. A single systematic shift is
mechanistically consistent with shared-die contention specifically;
scattered, bidirectional noise of comparable or greater magnitude is
not evidence against dedicated access being well-behaved, but it is
evidence that exclusive allocation does not, by itself, eliminate
run-to-run variation -- only its systematic, one-directional form.

The practical consequence is not that $\Gamma$ cannot be applied to
virtualized platforms, but that doing so requires treating $\rho_M$
itself as uncertain in a specific, identifiable way -- a systematic
offset, not merely elevated noise -- rather than uncertain in the same
generic sense dedicated hardware's own ordinary variance already is.
Predictions derived from a virtualized platform's nominal
specifications should be read with this in mind: a single measurement
disagreeing with a prediction is now weaker evidence against $\Gamma$
than the same disagreement on dedicated hardware would be, since a
one-off systematic shift, not scattered noise, is the specific,
already-observed failure mode on that class of platform; repeated
measurements agreeing with each other despite this risk remain the
stronger standard of evidence either way.

\subsection{A Corollary: Predictable Correlation Across Dedicated, Architecturally Diverse Machines}
\label{subsec:cross-architecture-corollary}

Section~\ref{subsec:rho-m-precondition} states what dedicated access
guarantees for a single machine: a fixed, well-defined $\rho_M$. A
stronger claim follows from it, worth stating explicitly rather than
leaving implicit in the accumulated results: on dedicated hardware,
$\Gamma$'s predicted parameters and a machine's known capacities and
bandwidths should stand in a predictable, derivable relationship
\emph{regardless of the specific architecture measured} -- CPU or GPU,
one vendor or another -- provided $\rho_M$ is accurately characterized
for that architecture's actual memory hierarchy. This is a claim about
correlation holding \emph{across} architecturally diverse dedicated
machines, not merely within one.

Delta already provides one confirmed instance: $M_C=256$, derived from
its real 512~KB L2 with no free parameters fit after the fact, beat
the M1-Pro-inherited $M_C=64$ by 30--59\% at every size tested
(Section~\ref{sec:recovering}). That result alone establishes the
claim for one architecture. The pending A100 prediction from this
paper's companion empirical study
(TILE\_SIZE\(\approx\)72, derived from A100's own
164~KB shared-memory-per-SM figure) is
the next, architecturally distinct test of the same corollary: a CPU
result and a GPU result, both dedicated, both derived from the
identical formula applied to each machine's own real, independently
confirmed capacity. Agreement on both would be evidence the
correlation genuinely tracks known hardware sizes and speeds across
architecture families, not something specific to cache-based CPU
designs; agreement on Delta's CPU result without a corresponding A100
GPU result would narrow the corollary's scope to architectures sharing
Delta's specific memory model, still a meaningful but more limited
finding.

\section{Recovering Known Results}
\label{sec:recovering}

$\Gamma$ is only useful if it reproduces what direct calibration
already found, on every machine measured to date, as a special case
rather than a coincidence restricted to the machine it was
reverse-engineered from.

\begin{table}[h]
\centering
\small
\caption{$\Gamma$ applied to four machines' shapes, against directly calibrated or independently derived values}
\label{tab:recovering}
\begin{tabular}{@{}p{0.19\textwidth}rp{0.27\textwidth}p{0.26\textwidth}@{}}
\toprule
\textbf{Machine} & \textbf{$\Gamma$-derived} & \textbf{Calibrated/derived independently} & \textbf{Source} \\
\midrule
M1 Pro, $M_C$ & 64 (\(\sigma=1\), L1) & 64 & \cite{Mullin2026GEMM}, retrospective \\
M1 Pro, $N_C$ & 2048 (\(\sigma=1/6\), L2) & 2048 & \cite{Mullin2026GEMM}, retrospective \\
Graviton4, $N_C$ (naive \(\sigma\) transfer) & 3072 & 512 (empirical) & mismatch, motivates this paper \\
A100, TILE\_SIZE & 72 (double-buffered) & --- & pending \\
\bottomrule
\end{tabular}
\end{table}

The first two rows are not independent confirmations -- they are the
values $\Gamma$'s occupancy fractions were set to reproduce, stated here
only to confirm the recursive machinery of
Section~\ref{sec:formalization} outputs them correctly given those
occupancy fractions. The third row is the paper's motivating failure, restated in
this notation: $\Gamma$ applied with M1-Pro-calibrated occupancy fractions
transferred to Graviton4's known capacities overpredicts by $6\times$.
The fourth row states a prediction not yet tested against measurement:
a generic double-buffering argument ($\sigma = 1/2$, motivated by
needing room for both the current and next tile simultaneously, not
reverse-engineered from any prior GPU measurement) applied to A100's
real, independently confirmed 164~KB shared-memory-per-SM figure
predicts TILE\_SIZE=72. This prediction is stated here before the
corresponding measurement is available, in keeping with the standard
the rest of this table and its
companion empirical study hold themselves to.

\section{The Open Problem: Is $\sigma_i$ Derivable, or Calibrated?}
\label{sec:open-problem}

Section~\ref{sec:recovering} isolates the paper's actual unresolved
question precisely: $\Gamma$'s recursive structure and its two
governing conditions (Equations~\ref{eq:capacity}
and~\ref{eq:intensity}) are not in dispute by the evidence gathered so
far, but $\sigma_i$ is a free parameter in that structure, and only
one of the four rows in Table~\ref{tab:recovering} used a
$\sigma_i$ that was not fit to a specific machine after the fact --
and that one row is still pending confirmation. We
state four candidate hypotheses for what determines $\sigma_i$,
each falsifiable independently, rather than treating the question as
open in only a vague sense.

\paragraph{H1: universal constant.} $\sigma_i$ takes a fixed value
for all private levels and a fixed (possibly different) value for all
shared levels, independent of architecture. This is already falsified
by the Graviton4 result in Table~\ref{tab:recovering}'s third row: the
same shared-level $\sigma=1/6$ that held for the M1 Pro's L2
overpredicted Graviton4's $N_C$ by $6\times$.

\paragraph{H2: co-tenancy-scaled.} For a shared level accessed by $P$
concurrent threads or cores, $\sigma_i \propto 1/P$: the occupancy fraction
shrinks in proportion to how many consumers must share the
level simultaneously. This is not yet falsified, but is not exactly
confirmed either: the M1 Pro's empirical $\sigma = 1/6$ at $P=8$ is
$1.33\times$ the naive $1/P = 1/8$ prediction, close enough to
suggest co-tenancy is part of the explanation and far enough from
exact agreement to suggest it is not the whole explanation. This
statement of H2 is itself ambiguous between two distinct readings,
tested separately in Section~\ref{subsec:specific-experiment}: (H2-raw)
$\sigma$ as a direct fraction of the level's total capacity, or
(H2-normalized) $\sigma$ as a fraction applied \emph{after} first
dividing capacity by $P$, i.e., against each thread's fair share of
the level rather than the level as a whole. The two readings turned
out to give different verdicts on real hardware.

\paragraph{H3: associativity- or prefetcher-dependent.} $\sigma_i$
depends on properties of the cache implementation not captured by
capacity, bandwidth, or thread count alone -- set-associativity
(which determines how much of a cache's nominal capacity is
practically reachable by a given access pattern before conflict
misses appear) or hardware prefetcher aggressiveness (which can
partially substitute for cache residency on chips with more capable
prefetchers, of which Apple Silicon is a documented example, making a
smaller occupancy fraction viable there than capacity arguments alone would
suggest). Neither property is reliably published by vendors at the
level of detail this hypothesis would need to test directly; testing
H3 would require inferring these properties indirectly, from
architectures with documented differences along one axis while
holding others as fixed as possible.

\paragraph{H4: private-level-count-dependent.} $\sigma_i$ for a shared
level depends on how many private cache levels sit between a thread
and that shared level, independent of co-tenancy. The M1 Pro's shared
L2 is reached through a single private level (L1 only); Delta's and
Graviton4's shared levels are each reached through two (L1 and a
second private level -- L2 on both, notably large on Graviton4 at
2~MB). Compared on equal footing -- raw $\sigma$, not mixing raw and
per-thread-normalized readings -- the direction is consistent across
all three machines measured to date: M1 Pro's single-private-level
$\sigma=1/6$ is roughly $5$--$10\times$ larger than either
two-private-level machine's ($1/36$ for Graviton4, $1/64$ for Delta).
The two two-private-level machines do not match each other exactly --
a further $\sim\!2\times$ gap remains between them -- so H4 does not
explain $\sigma$ completely on its own, but the qualitative pattern
(more private absorption before the shared level, smaller occupancy fraction
needed there) holds directionally across every machine tested so far,
not merely the two that motivated H2's normalized reading. H4 and H2
are not competing explanations so much as different axes of the same
underlying question; a machine with matched private-level count but
different co-tenancy, or matched co-tenancy but different
private-level count, would be needed to cleanly separate them.

\subsection{A Specific, Falsifiable Experiment}
\label{subsec:specific-experiment}

Delta CPU's per-CCD structure (Section~\ref{sec:recovering}'s source
study; 8 cores sharing a 32~MB L3 slice) makes a controlled test of H2
directly available: restricting the fixed-$N_C$ sweep to exactly one
CCD (8 threads, pinned via \texttt{taskset} to CPUs sharing a single
L3 group, confirmed via \texttt{lscpu -e}) and sweeping $N_C$ against
that CCD's 32~MB L3 reproduces the M1 Pro's $P=8$ co-tenancy count on
a different chip, cache capacity, and microarchitecture entirely.

\textbf{This experiment has since been run.} Pinned to one CCD, at
the corrected $M_C=256$ (Section~\ref{sec:recovering}), the empirical
optimum is $N_C=256$, decisively and consistently across all four
matrix sizes tested (4--11\% ahead of the next-best value at every
size, a substantially cleaner signal than any unpinned sweep
produced). The two readings of H2 give sharply different verdicts on
this result:

\begin{itemize}
\item \textbf{H2-raw is falsified.} $\sigma = (N_C \cdot K_C \cdot
8\text{ bytes}) / 32\text{MB} = 1/64$, a $10.7\times$ mismatch against
the M1 Pro's $1/6$, despite $P=8$ matching exactly on both machines.
Co-tenancy count alone, applied as a direct fraction of total
capacity, does not determine $\sigma$.

\item \textbf{H2-normalized survives.} Dividing the CCD's 32~MB by
$P=8$ first (each thread's nominal fair share, 4~MB) and computing
$\sigma$ against that share instead gives $\sigma' = 1/8$ --
$0.75\times$ the M1 Pro's $1/6$, the same rough magnitude of
discrepancy the M1 Pro's own empirical value already showed against
naive $1/P$ before this experiment was run. Co-tenancy, correctly
normalized, is not falsified by this result.
\end{itemize}

\begin{figure}[h]
\centering
\includegraphics[width=0.85\textwidth]{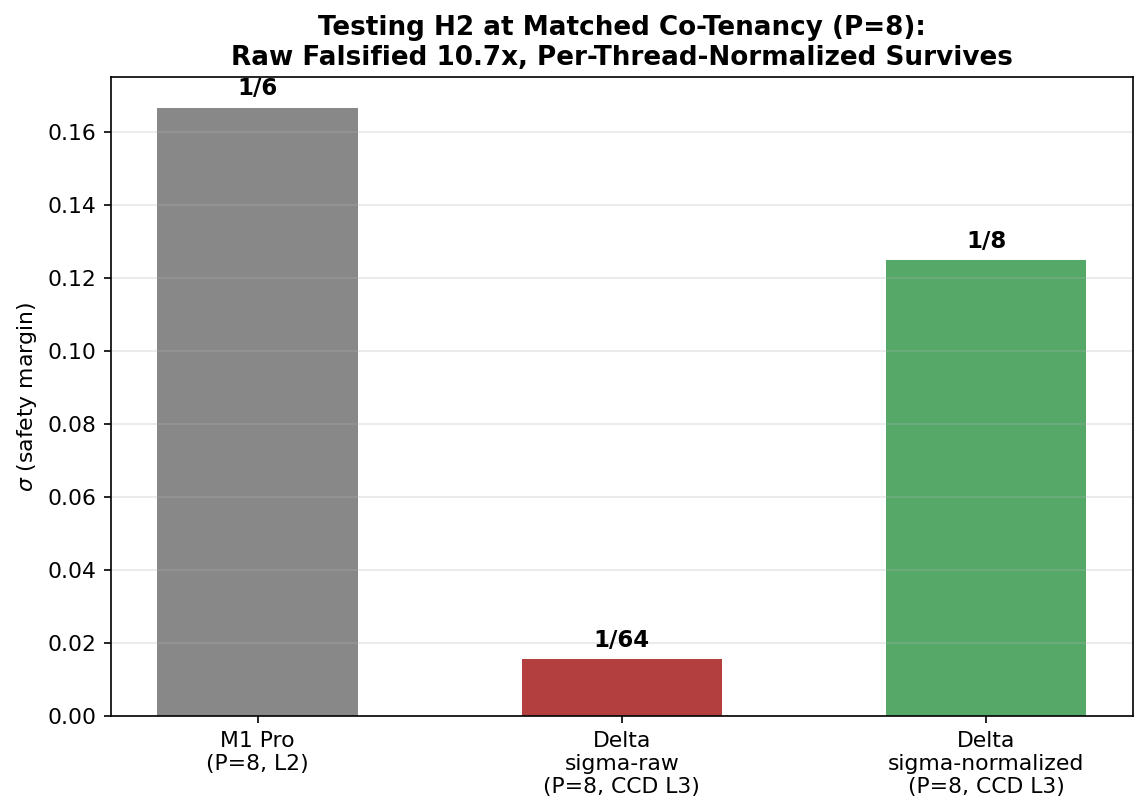}
\caption{The two readings of H2, tested at matched co-tenancy (P=8) on Delta against the M1 Pro's own value. The raw reading (red) is off by an order of magnitude; the per-thread-normalized reading (green) lands in the same range as the M1 Pro (gray).}
\label{fig:delta-sigma-h2}
\end{figure}

The honest reading is that H2 was underspecified rather than simply
right or wrong: the ambiguity between these two formulations was not
resolved by the M1 Pro data alone, and this experiment is what
distinguishes them. Both readings still leave a residual, unexplained
gap ($10.7\times$ for the raw form's failure mode; $\sim\!1.3\times$
for the normalized form's partial success). H4
(Section~\ref{sec:open-problem}) accounts
for part of this directionally -- Delta's two private levels against
the M1 Pro's one is consistent with Delta's smaller raw $\sigma$ --
without explaining it completely, since Graviton4, with the same
private-level count as Delta, does not match Delta's value either.
H3's candidates -- Milan's L3 associativity, or its prefetcher
behavior relative to Apple Silicon's -- remain live for whatever H4
does not account for, rather than co-tenancy or private-level count
alone explaining $\sigma$ completely under any reading tested so far.

\section{From Registers Outward: A More Principled Base Case, and the Limits of Extending to a Network}
\label{sec:inside-out}

Section~\ref{sec:formalization} defined $\Gamma$ recursively from the
outermost level (DRAM or HBM) inward, terminating arbitrarily when
$\rho_M$ is exhausted. This section reformulates the same recursion in
the opposite direction, shows the reformulation changes the
justification without changing any derived value, and uses that
clearer justification to state precisely how far the framework
extends toward a network level -- and exactly where it stops
extending cleanly.

\subsection{Reformulation: Registers as a True Base Case}
\label{subsec:inside-out-reformulation}

The register-level tile, $M_R \times N_R$, is not chosen by any
capacity or bandwidth argument -- it is fixed by the instruction set
architecture: vector width and the number of architectural registers
available to hold accumulators. Every level outward from it exists,
physically, to keep the level just inside it fed continuously,
without stalling. Restating $\Gamma$ in this direction: level $0$
(registers) is the unconditioned base case, and for each level $i+1$
moving outward, $\tau_{i+1}$ is chosen so that (a) it satisfies
Equation~\ref{eq:capacity} against $\lambda_{i+1}$, and (b) the time
spent computing on level $i$'s tiles, drawn from a resident
$\tau_{i+1}$, is at least the time required to fetch the next
$\tau_{i+1}$-sized tile from level $i+2$ -- the same arithmetic as
Equation~\ref{eq:intensity}, now justified as ``level $i+1$ must not
itself stall level $i$'s consumption'' rather than ``level $i+1$ must
fit within an externally imposed budget.''

These two formulations compute identical values: applied to the M1
Pro's two-level hierarchy, both give $M_C=64$ and $N_C=2048$, since
the underlying equations are unchanged. What changes is which
direction is physically motivated. Outward-in leaves the recursion's
termination -- why registers need no governing equation -- unexplained
except by fiat. Inside-out makes registers a true base case (fixed by
the ISA, prior to any capacity or bandwidth reasoning) and makes the
prefetch policy of Section~\ref{sec:formalization} a direct
consequence of keeping a specific, named consumer fed, rather than an
abstract latency-hiding argument stated separately from the capacity
condition it accompanies.

\subsection{Extending Outward: A Network Level, and Where the Extension Stops Being Free}
\label{subsec:network-extension}

Stated inside-out, the natural question is how far the recursion
extends past DRAM. Treating a distributed-memory network as one
further level, $\lambda_{net} = (C_{net}, B_{net}, \sigma_{net})$,
with DRAM playing the role of ``the level being kept fed,'' the same
two conditions restate without new machinery:
\[
\text{size}(\tau_{net}) \leq \sigma_{net} \cdot C_{net}, \qquad
\frac{\text{FLOPs}(\tau_{net})}{\text{size}(\tau_{net})} \geq
\frac{\text{PeakFLOPrate}}{B_{net}}
\]
where $C_{net}$ is, concretely, local DRAM capacity (reduced to account for
whatever else must reside there simultaneously, including the packed
buffers for every cache level beneath it), and $\tau_{net}$ is how
much of the global problem one node holds locally.

This is where the extension stops being free, and stating the limit
precisely is more useful than eliding it. Every cache-level transition
in Section~\ref{sec:formalization} is point-to-point: level $i$
requests data from level $i{+}1$, receives it, and $B_i$ is a fixed
physical property of that pair of levels, independent of what
algorithm is being executed. Realizing $\tau_{net}$ across multiple
nodes is not point-to-point in the same sense -- it requires an
explicit distributed algorithm (SUMMA, Cannon's algorithm, or a 2.5D
variant, among others) specifying which node sends what to whom, and
different algorithms realizing the identical $\tau_{net}$ achieve
different effective bandwidths on the identical physical network,
because collective communication patterns (broadcasts, shifts,
all-to-all exchanges) have their own, topology-dependent scaling
behavior that a single point-to-point $B_{net}$ cannot capture.
Concretely: $B_{net}$ in the equations above is not a fixed hardware
constant the way $B_{L1}$, $B_{L2}$, and $B_{L3}$ are -- it is a
function of which communication algorithm is chosen, and $\Gamma$ as
defined does not choose that algorithm. The capacity and
latency-hiding conditions correctly determine \emph{how much} data a
node should hold; they do not, without an additional and separate
choice of communication pattern, determine \emph{how} that data
should move between nodes to realize the tiling they prescribe.

\subsection{A Nearer-Term Test: NUMA as a Miniature Network}
\label{subsec:numa-miniature}

A true multi-node cluster is not the only, or the nearest, place this
question can be tested. Delta CPU's dual-socket topology
(Section~\ref{sec:recovering}'s source study) already contains a
smaller version of the same phenomenon: a thread accessing the other
socket's DRAM crosses AMD's Infinity Fabric, a physically distinct
path from local DDR4 channels, with its own bandwidth and latency
characteristics. Two sockets are too few to exhibit genuine collective
communication -- there is only one other node to address, not many --
but the core claim of Section~\ref{subsec:network-extension} does not
require collectives to be tested: it requires only that
\emph{achieved} bandwidth depend on \emph{how} data is placed and
accessed, not solely on the physical link's rated capacity.

This experiment has since been run. Thread count was held fixed at 64
throughout, so that only memory placement varied between two
conditions, forced via \texttt{numactl}: \emph{local}
(compute and memory both bound to one socket, zero cross-socket
traffic by construction) against \emph{remote} (compute bound to the
second socket while memory remains forced to the first, so every
single access crosses the socket boundary). Raw STREAM bandwidth
differed by under 1\% between the two conditions -- if bandwidth
capacity alone determined achieved performance, cross-socket access
would look almost free. The matrix multiplication kernels measured
alongside it did not agree: GEBP lost 13.9\% and a recursive-classical
kernel 15.1\% under the remote condition, while an MoA-pipelined
kernel lost only 2.4\%, a fivefold smaller penalty under the identical
remote-memory condition. This is direct, measured evidence for
Section~\ref{subsec:network-extension}'s claim, not merely
smaller-scale support for it: access pattern, not physical bandwidth
alone, determines what a level actually delivers, and different
kernels computing the identical mathematical result are not equally
exposed to that difference. The companion empirical study reports
this measurement in full \cite{Mullin2026DeltaResults}; this section
draws the conclusion for the framework specifically that MoA-pipelined's
own access pattern is measurably more resilient to non-local memory
access than GEBP's blocked-and-packed pattern, on this specific
hardware -- a property neither this paper nor the conjecture it tests
predicted in advance, and worth stating plainly as a genuine finding
rather than a confirmation of something already expected.

\section{Toward a Predictive Performance Equation}
\label{sec:predictive-equation}

Everything so far predicts \emph{parameters} -- $M_C$, $N_C$, a tile
size. This section states, as precisely as the evidence allows, what
it would take to predict \emph{throughput itself} on a machine not yet
tested, and reports a first, honest attempt.

\subsection{The Equation, and Which Terms Are Which}
\label{subsec:equation-terms}

The natural extension of a roofline model, with $\Gamma$ supplying the
blocking rather than leaving it as a free parameter, is
\[
\text{GFLOPS}(n, P) = \min\Big(P \cdot \text{achieved}(1) \cdot \eta(P),\ \ AI(M_C,K_C,N_C) \cdot B_{\text{raw}} \cdot \rho_{\text{access}}(\text{kernel})\Big)
\]
with $(M_C, K_C, N_C)$ given by $\Gamma(\rho_M)$, not fit to this
equation separately. Each remaining term falls into one of three
categories, worth distinguishing sharply rather than treating the
equation as uniformly known or uniformly speculative:

\begin{itemize}
\item \textbf{Derivable from a specification sheet, no measurement
required:} $M_C$, $N_C$ (Section~\ref{sec:formalization}), and
$AI(M_C,K_C,N_C)$, the standard arithmetic-intensity formula for a
blocked GEBP-style kernel.
\item \textbf{Requires exactly one real measurement, not derivable
from a spec sheet:} $\text{achieved}(1)$, single-thread throughput on
the actual compiled binary. An early attempt at this equation used
each machine's theoretical peak FLOP rate (vector width times clock)
in place of a measured baseline, and overshot Delta's own measured
$P=8$ throughput by $6.6\times$ -- not because parallel efficiency is
that poor, but because theoretical peak assumes vectorization no
plain, portably-compiled C code achieves. $\eta(P)$ computed against a
measured baseline recovers exactly the values already reported for
Delta's own contention sweep in the companion empirical study
\cite{Mullin2026DeltaResults}; computed against theoretical peak, it
silently absorbs a confound that has nothing to do with parallelism.
The corrected equation requires the real measurement precisely because
skipping it produces a number that is wrong for a specific, identified
reason, not merely imprecise.
\item \textbf{Not yet reducible to a formula at all, only to discrete
measured points:} $\eta(P)$ itself, and $\rho_{\text{access}}$.
\end{itemize}

\subsection{Testing Transferability: Does $\eta(P)$ Have One Shape?}
\label{subsec:eta-transferability}

The most useful question about $\eta(P)$ is not its value on any one
machine but whether its \emph{shape}, once normalized against each
machine's own single-thread baseline, is shared across machines --
which would make it a transferable term rather than one requiring
fresh calibration every time. This is directly testable with data
already in hand: identical calibration sweeps exist for M1 Pro,
Graviton4, and Delta, all at $P=1,2,4,8$, all normalizable the same
way.

\begin{figure}[h]
\centering
\includegraphics[width=0.85\textwidth]{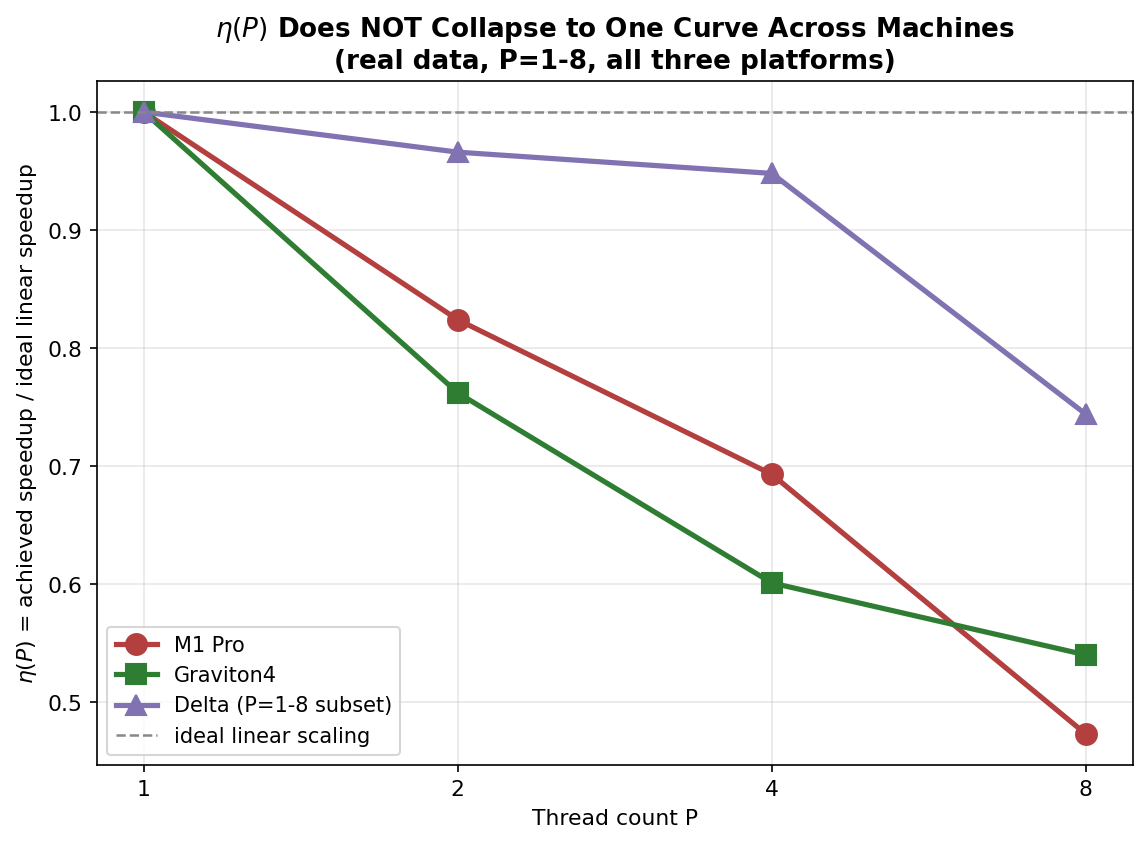}
\caption{$\eta(P)$ computed identically (measured throughput, normalized to each machine's own $P{=}1$ baseline) for all three platforms tested to date.}
\label{fig:eta-three-machines}
\end{figure}

It does not collapse to one curve. Delta stays markedly closer to
ideal ($\eta(8)=0.744$) than M1 Pro ($0.473$) or Graviton4 ($0.540$),
which more closely track each other. This is a real, quantified
answer, not an absence of one: $\eta(P)$ requires genuine per-machine
calibration under current understanding, and the equation above cannot
be applied to an untested machine using another machine's $\eta(P)$
values, even a topologically similar one. A tempting explanation --
that $\eta(P)$'s decline tracks shared cache size per core -- does not
hold up against just these three points: Graviton4's shared L2
(36~MB) is larger than M1 Pro's (24~MB) yet shows a similar decline,
not a shallower one. Three machines is not enough to identify what
does drive the difference; it is enough to rule out the first,
simplest guess.

\subsection{A Sharper Comparison: Efficiency at Each Machine's Own Full Utilization}
\label{subsec:full-utilization}

Comparing $\eta(P)$ at a fixed $P=8$ across machines, as above,
compares quantities that mean different things per machine: $P=8$ is
the entire M1 Pro and the entire Graviton4 instance, but one-sixteenth
of Delta's 128-core node. A more meaningful comparison holds not $P$
fixed but the \emph{fraction of each machine's own total capacity}
fixed, comparing every machine at its own $P_{\max}$:
\[
\eta(P_{\max}) = \frac{\text{achieved}(P_{\max})}{P_{\max} \cdot \text{achieved}(1)}
\]
\begin{center}
\begin{tabular}{lrr}
\toprule
Machine & $P_{\max}$ & $\eta(P_{\max})$ \\
\midrule
M1 Pro & 8 & 0.473 \\
Graviton4 & 8 & 0.540 \\
Delta & 128 & 0.150 \\
\bottomrule
\end{tabular}
\end{center}
This reverses the ordering the fixed-$P=8$ comparison gave in
Section~\ref{subsec:eta-transferability}, where Delta appeared to
scale best. Using the entirety of a 128-core node costs substantially
more relative efficiency than using the entirety of an 8-core one --
a result that aligns with, rather than against, the original
conjecture's underlying concern that higher core counts would show
more pronounced overhead, even though the specific mechanism
predicted (fork/join cost scaling with thread count within a fixed
core budget) is not the same claim as this one (efficiency measured
against each machine's own ceiling). The two fixed-$P$ and
full-utilization comparisons are not in conflict; they answer
different questions, and neither alone is sufficient to characterize
$\eta$ across machines of different scale.

\subsection{What This Section Does and Does Not Establish}
\label{subsec:predictive-status}

This is not a working predictive model for an arbitrary new machine.
It is a precise accounting of what such a model would require: two
terms derivable today from nothing but a specification sheet, one term
requiring a single cheap measurement rather than a full experimental
campaign, and one term -- the one this section spent most of its
effort on -- shown, honestly, not to transfer between machines under
the simplest hypothesis tested. Stating this precisely is more useful
than either overclaiming a general formula or abandoning the attempt:
the next machine tested will extend $\eta(P)$ from three points to
four, and either begin to suggest what actually governs it, or rule
out a second hypothesis the way shared-cache-size-per-core was just
ruled out.

\section{Conclusion}
\label{sec:conclusion}

This paper formalized a question the companion empirical study raised
but did not itself resolve: whether known memory-hierarchy sizes can
predetermine blocking and prefetching, or only after empirical
calibration. The answer, made precise rather than left as a general
impression, is split cleanly by the formalization itself. The
recursive structure -- $\Gamma$, derived from a machine shape
$\rho_M$ the same way MoA's existing $\gamma$ is derived from an
array shape $\rho A$, governed by a capacity condition and a
latency-hiding condition applied once per level -- appears to transfer
correctly across every architecture examined so far, with one
prediction pending confirmation: a double-buffered tile bound whose
occupancy fraction was derived from architecture-independent
reasoning rather than another machine's fitted value, stated before
measurement in the same spirit as every other conjecture in this
paper. Restating the same recursion inside-out, from
a register-level base case fixed by the instruction set architecture
rather than an arbitrary termination, changes no derived value but
makes precise exactly how far this reasoning extends: cleanly through
every cache level, and only partway to a distributed-memory network,
where the equations correctly determine how much data a node should
hold but not, without an additional and separate choice of
communication algorithm, how that data should move between nodes to
realize it -- a genuine limit of the framework, stated as such rather
than elided. A smaller-scale test of the same underlying claim, run on
Delta's own dual-socket topology rather than requiring a multi-node
cluster, confirmed it directly: holding thread count fixed and varying
only memory placement, kernels computing the identical mathematical
result were not equally exposed to non-local memory access, a
fivefold difference in penalty between two kernels under the identical
physical condition. The occupancy fractions $\sigma_i$
do not transfer across architectures in their naive form: a
co-tenancy-matched experiment on real Delta hardware falsified the
direct reading of the leading hypothesis decisively, while the same
hypothesis, correctly normalized per thread rather than against total
capacity, survived with a residual gap comparable to what motivated it
on the M1 Pro in the first place. This is not a full resolution --
associativity and prefetcher-dependence remain live candidates for
that residual gap -- but it is a real result rather than a stated
intention: the ambiguity between two readings of the same hypothesis
was not resolvable from the M1 Pro's data alone, and required a
second, sufficiently different machine to expose. A further corollary
follows from the dedicated-access precondition stated alongside
$\Gamma$'s definition: on hardware where $\rho_M$ is genuinely
well-defined, the correlation between known sizes and speeds and
$\Gamma$'s predicted parameters should hold across architecturally
distinct machines, not merely within one. Delta's confirmed $M_C$
result establishes this for one architecture; a pending, identically
derived prediction on A100 -- a genuinely different architecture
sharing only the property of dedicated access -- is the next test of
whether that correlation is a property of cache-based design in
general or was specific to the CPU architectures examined so far.

\section*{Acknowledgments}

This work used Delta at the National Center for Supercomputing
Applications and Anvil at Purdue University through allocations
bibg-delta-cpu and CIS261396 from the Advanced Cyberinfrastructure
Coordination Ecosystem: Services \& Support (ACCESS) program, which is
supported by National Science Foundation grants \#2138259, \#2138286,
\#2138307, \#2137603, and \#2138296 \cite{Boerner2023}. The Delta
advanced computing resource is a collaborative effort between the
University of Illinois Urbana-Champaign and its National Center for
Supercomputing Applications, supported by the National Science
Foundation (award OAC 2005572) and the State of Illinois. Anvil is
funded under National Science Foundation award No.\ 2005632.

\bibliographystyle{plain}
\bibliography{references}

@article{Goto2008,
  author  = {Kazushige Goto and Robert A. van de Geijn},
  title   = {Anatomy of High-Performance Matrix Multiplication},
  journal = {ACM Transactions on Mathematical Software},
  volume  = {34},
  number  = {3},
  pages   = {1--25},
  year    = {2008},
  doi     = {10.1145/1356052.1356053}
}

@phdthesis{Mullin1988,
  author  = {Lenore M. Mullin},
  title   = {A Mathematics of Arrays},
  school  = {Syracuse University},
  address = {Syracuse, NY, USA},
  year    = {1988},
  type    = {{Ph.D.} dissertation}
}

@unpublished{Mullin2026GEMM,
  author = {Lenore M. Mullin},
  title  = {Mathematics of Arrays for High-Performance Dense Matrix Multiplication: An Experimental Comparison of Blocking and Strassen Using Roofline Analysis},
  note   = {Submitted to ACM Transactions on Mathematical Software},
  year   = {2026}
}

@misc{Mullin2023GPU,
  author = {Lenore M. R. Mullin},
  title  = {From Array Algebra to Energy Efficiency on GPUs: Data and Hardware Shapes with Dimension-Lifting to Optimize Memory-Processor Layouts},
  note   = {arXiv:2306.11148},
  year   = {2023}
}

@unpublished{Mullin2026DeltaResults,
  author = {Lenore M. Mullin},
  title  = {Testing the EPYC Conjecture on Real Hardware: MoA-Guided Dense Matrix Multiplication on NCSA Delta (AMD EPYC 7763 "Milan")},
  note   = {Companion empirical results paper},
  year   = {2026}
}

@inproceedings{Boerner2023,
  author = {Timothy J. Boerner and Stephen Deems and Thomas R. Furlani and Shelley L. Knuth and John Towns},
  title  = {ACCESS: Advancing Innovation: NSF's Advanced Cyberinfrastructure Coordination Ecosystem: Services \& Support},
  booktitle = {Practice and Experience in Advanced Research Computing (PEARC '23)},
  year   = {2023},
  address = {Portland, OR, USA},
  publisher = {ACM},
  doi    = {10.1145/3569951.3597559}
}

\end{document}